\documentclass[twocolumn,nofootinbib,amsmath,amssymb,aps,prd,balancelastpage,superscriptaddress]{revtex4-1}

\usepackage{color}
\usepackage[dvipsnames]{xcolor}
\usepackage[active]{srcltx}
\usepackage{amsmath,amsfonts,amssymb,amsthm,amstext,amscd,eucal,srcltx}
\usepackage{epsfig,graphicx,bm}
\usepackage{epstopdf, epsf}
\usepackage{dcolumn}
\usepackage{hyperref}
\usepackage{tensor}
\usepackage{lipsum}
\usepackage{appendix}
\usepackage{tikz}
\usepackage{caption}
\usepackage{subcaption}

\newcommand{\be}{\begin{equation}}
\newcommand{\ee}{\end{equation}}

\newcommand{\bse}{\begin{subequations}}
\newcommand{\ese}{\end{subequations}}
\newcommand{\bea}{\begin{eqnarray}}
\newcommand{\eea}{\end{eqnarray}}
\newcommand{\ba}{\begin{array}}
\newcommand{\ea}{\end{array}}
\newcommand{\bc}{\begin{center}}
\newcommand{\ec}{\end{center}}

\begin{document}

\vspace*{3mm}

\title{Gravitational Waves from Inflaton Decay and Bremsstrahlung}

\author{Anna Tokareva}
\email{tokareva@ucas.ac.cn}
\affiliation{School of Fundamental Physics and Mathematical Sciences, Hangzhou Institute for Advanced Study, UCAS, Hangzhou 310024, China}
\affiliation{International Centre for Theoretical Physics Asia-Pacific, Beijing/Hangzhou, China}
\affiliation{Theoretical Physics, Blackett Laboratory, Imperial College London, SW7 2AZ London, U.K.}

\begin{abstract}
\noindent
The concept of early Universe inflation resolves several problems of hot Big Bang theory and quantitatively explains the origin of the inhomogeneities in the present Universe. However, it is not possible to arrange inflation in a scalar field model with renormalizable potential, such that it would not contradict the recent Planck data. For this reason, inflaton must have also higher derivative couplings suppressed at least by the Planck scale. We show that these couplings may be relevant during reheating and lead to non-negligible production of gravitons. We consider the possibility that the unitarity breaking scale for the model of inflation is lower than the Planck scale and compute production of gravitons during reheating, due to the inflaton decay to two gravitons and graviton bremsstrahlung process. The spectrum of produced gravitons is crucially dependent on reheating temperature and inflaton mass. We find that for low reheating temperature decay to gravitons lead to significant amount of dark radiation. Confronting this result with CMB constraints, we find reheating dependent bounds on the unitarity breaking scale. We also compare the obtained gravitational wave signals with the projected limits of future high frequency gravitational wave experiments.
\end{abstract}

\thanks{Preprint: Imperial/TP/2023/AAT/1}

\maketitle

\section{Introduction}

The stage of exponential expansion preceding the hot Big Bang serves as a good description of the origin of cosmological perturbations \cite{Starobinsky:1980te,Linde:1981mu,Mukhanov:1981xt} (see also a review \cite{Martin:2013tda} and references therein), consistent with all the existing data. The simplest scenarios assume that it was driven by a single scalar field slowly rolling towards the minimum of its potential,
\begin{equation}
    S=\int d^4 x \sqrt{-g}\left(\frac{M_P^2}{2}R-\frac{1}{2}\partial_{\mu}\phi \partial^{\mu}\phi-V(\phi)\right).
\end{equation}

As this theory contains gravity, it becomes non-predictive around the Planck scale. It can happen even at lower scale, depending on the choice of inflaton potential (the well-known example is Higgs inflation \cite{Bezrukov:2007ep}, the unitarity issue is discussed in \cite{Bezrukov:2010jz}). Thus, the relevant description of inflation should be an effective field theory (EFT) which is valid only below some scale (we refer to it as $\Lambda_{UV}$ -- unitarity breaking scale or cutoff scale. This scale can be estimated from the EFT expansion as the lowest energy scale suppressing non-renormalisable operators \cite{Bezrukov:2010jz}. More precisely, this scale can be obtained from the partial wave expansion of scattering amplitudes  and requirement of unitarity \cite{0906.2349} (see aslo \cite{Ageeva:2022fyq, Ageeva:2022nbw} for the recent applications to cosmology).

We consider here the case when both inflation and reheating, as well as transition between these stages (preheating), can be described by means of perturbation theory in an EFT withing the regime of its validity. We keep the choice of inflaton potential arbitrary, as our results will be dependent only on its expansion around the flat space minimum, i.e. on the inflaton mass. Many phenomenological models suggested for describing inflation have a EFT unitarity breaking scale $\Lambda_{UV}$ lower than the Planck mass, see for example \cite{Bauer:2010jg,Shaposhnikov:2020gts,He:2023fko,Langvik:2020nrs,Kallosh:2013hoa,Galante:2014ifa,Koshelev:2022olc,Koshelev:2023elc}. Also, the scale of quantum gravity may be lower than Planck mass due to loop contributions of a large number $N$ of matter fields, $\Lambda_{UV}\sim M_P/\sqrt{N}$ \cite{0706.2050,0710.4344,0806.3801,0812.1940}. 

However, these models can still be a valid description of inflation if the Hubble scale of inflation $H_{inf}$ is much lower than $\Lambda_{UV}$. They can also describe preheating and reheating if production of matter particles is slow and inflaton mass $m\ll \Lambda_{UV}$. In this case reheating can be described as a perturbative decay of the oscillating inflaton condensate. The latter supports early matter-dominated stage which lasts until the inflaton decay width to the Standard Model (SM) particles becomes of order of the Hubble scale $\Gamma_{SM}\sim 2/(3 t)$. Transition to the SM radiation dominated stage happens around that time.

In this work we make the following assumptions about the cosmological model of inflation and reheating:
\begin{itemize}
    \item Inflation was driven by a single field $\phi$ with a potential $V(\phi)$ providing the Hubble parameter $H_{inf}$.
    \item Inflation ended by oscillations of the inflaton field around its minimum $\phi=0$, where $V(\phi)\approx m^2\phi^2/2$.
    \item This field decays to the SM particles leading to reheating temperature $T_{reh}$.
    \item The unitarity breaking scale $\Lambda_{UV}$ of EFT describing both inflation and reheating is between $H_{inf}$ and $M_P$. We also require $\Lambda_{UV}\gg m$, so that the inflaton decay can be consistently described within the EFT.
\end{itemize}

We consider gravitational wave (GW) production in EFT of inflaton, matter and gravity which happens due to inevitable presence of higher derivative couplings between inflaton and gravity, as well as between matter and gravity. Gravitons can be produced due to the direct decay of inflaton to two gravitons and due to graviton bremsstrahlung when inflaton decays to two matter particles and one graviton. We study the relevance of both effects for high frequency GW signal and their contribution to the dark radiation which is severely constrained by CMB \cite{Planck:2018vyg}. As the main result of this work, we show the importance of non-renormalizable operators for graviton production in the early Universe. Their effect can be non-negligible even if they are suppressed by Planck scale.

The paper is organised as follows. In the next Section we introduce couplings between inflaton and gravity arising in a generic EFT, compute inflaton decay rate to gravitons and derive a lower bound on the reheating temperature from the Planck constraint on dark radiation. In Section 3, we consider the lowest order couplings between matter, inflaton and gravity and present the results for graviton bremsstrahlung rate. In Section 4 we compute the present day GW spectrum emerging from inflaton decay to two gravitons and from graviton bremsstrahlung during reheating. In Section 5 we discuss the obtained results and compare them with projected sensitivities of future GW experiments.

\section{EFT of inflaton and gravity: decay to gravitons}

In the context of EFT we consider here the following extra couplings between the inflaton and gravity appearing in the generic non-renormalizable model\footnote{We don't write the terms $\phi R$, $(\partial_{\mu}\phi)^2 R$ and $\phi R^2$ assuming that the theory was already brought to the Einstein frame by the proper metric field redefinition (Weyl transformation). Thus, we assume they are taken into account in the form of potential $V(\phi)$ and in the other inflaton self-couplings.},

\begin{widetext}
\begin{equation}
S_{NR}=\int d^4 x \sqrt{-g}\left(\frac{\phi}{\Lambda_1}R_{\mu\nu\lambda\rho}R^{\mu\nu\lambda\rho}+\frac{\phi}{\Lambda_2}R_{\mu\nu}R^{\mu\nu}+\frac{\phi}{\Lambda_3}R^2+\frac{1}{\Lambda_4^2}G_{\mu\nu}\partial^{\mu}\phi \partial^{\nu}\phi\right).
\end{equation}
\end{widetext}

The other couplings will be suppressed by even higher powers of the scales $\Lambda$. The expected hierarchy of these scales, in relation to $\Lambda_{UV}$, is $\Lambda_{1,2,3}\sim \Lambda_{UV}^3/M_P^2$\footnote{As it was recently claimed in \cite{Serra:2022pzl}, there is a very non-trivial causality bound $\Lambda_{UV}\lesssim \sqrt{\Lambda_1 M_P \log{(b_{IR}\Lambda_1)}}$, which, however, contains the IR impact parameter scale $b_{IR}$ and, thus, is eliminated in the formal limit of flat space $b_{IR}\rightarrow \infty$. In what follows, we refer to the naive power counting hierarchy, leaving the relevance of the mentioned bound in the cosmological context for future study.}, while $\Lambda_4\sim \sqrt{\Lambda_{UV}^3/M_P}$, see also \cite{1711.08761,Ruhdorfer:2019qmk}. Inflaton decay to two gravitons receives a contribution only from the term suppressed by $\Lambda_1$. The contribution from the other terms is vanishing on shell because external gravitons are transverse and traceless. The matrix element and decay width coming from the remaining term are ($k=m/2$),
\begin{equation}
   {i\cal M_{++}}={i\cal M_{--}}=\frac{8 k^4}{\Lambda_1 M_P^2},\qquad \Gamma_{GW}=\frac{m^7}{64\pi M_p^4\Lambda_1^2}
\end{equation}
Here $++$ and $--$ stand for the polarizations of the outgoing gravitons, ${\cal M_{+-}}={\cal M_{-+}}=0$. \footnote{These results, as well as the results for bremsstrahlung rate, were obtained with the use of xAct package of Wolfram Mathematica and substitution of the explicit form of graviton polarization tensors (CTensor package).} 

If inflaton is a pseudoscalar, in a CP-invariant theory it can have only a coupling to the gravitational Chern-Simons term leading to its decay to gravitons,

\begin{equation}
    L_{int}=\frac{\phi}{\Lambda_a}\epsilon_{\alpha\beta\gamma\delta}R^{\alpha\beta\mu\nu}R^{\alpha\beta}_{~~\,\mu\nu}.
\end{equation}
The decay rate following from this Lagrangian is
\begin{equation}
   {i\cal M_{++}}=-{i\cal M_{--}}=\frac{16 k^4}{\Lambda_a M_P^2},\qquad \Gamma_{GW}^a=\frac{m^7}{16\pi M_p^4\Lambda_a^2}
\end{equation}

If inflaton decay to matter particles described by the total decay rate $\Gamma_{SM}$ is responsible for reheating, the graviton contribution to the dark radiation (in terms of extra relativistic degrees of freedon $\Delta N_{eff}$) can be obtained in a standard way as \cite{1307.5298}
\begin{equation}
    \Delta N_{eff}=2.85 \frac{\rho_{GW}}{\rho_{SM}}=2.85\frac{\Gamma_{GW}}{\Gamma_{SM}}.
\end{equation}
One can see that this effect gets stronger for low reheating temperatures. In fact, setting a conservative requirement $\Delta N_{eff}\lesssim 0.2$ we obtain a lower bound on reheating temperature depending on the inflaton mass and UV cutoff scale,
\begin{equation}
    T_{reh}\gtrsim 0.15\, g_{reh}^{1/4}\frac{m^{7/2}}{M_P^{3/2}\Lambda_1}\left(\frac{\Delta N_{eff}}{0.2}\right)^{-1/2}.
\end{equation}
Remarkably, this bound is non-trivial and certainly should be taken into account even for Planck-suppressed couplings. For example, it gives numerically,
\begin{equation}
\begin{split}
   &m=10^{13}~{\rm GeV}\quad \rightarrow~ T_{reh}^{min}=1 ~{\rm GeV},\\
   &m=10^{16}~{\rm GeV}\quad \rightarrow ~T_{reh}^{min}=10^{10} ~{\rm GeV}.
   \end{split}
\end{equation}

\section{Couplings to matter: reheating and graviton bremsstrahlung.}

The transition of inflation to the Hot Big Bang is a model-dependent stage which is very hard to probe in observations. The maximal temperature of the thermalised SM radiation (reheating temperature) is bounded from BBN, $T_{reh}>1$ GeV \cite{Ringwald:2020ist}. The upper bound corresponds to the instant reheating when the energy of inflaton is transferred to the SM radiation right at the end of inflation ($T_{max}\sim 10^{15}$ GeV \cite{Ringwald:2020ist}). As we will show later, the gravitational wave spectrum crucially depends on the reheating temperature, as well as on the other details of reheating.

Here we specify the simplest reheating scenario implementing it by introducing the coupling between the inflaton and Higgs boson, 
\begin{equation}
\label{reh}
\begin{split}
S_{int}^{SM}=
\int d^4 x \sqrt{-g}\left(-|D_{\mu}H|^2+\mu \phi H^{\dagger}H +\right. \\
\left. -\frac{1}{\Lambda_5^2}G_{\mu\nu}D^{\mu}H^{\dagger}D^{\nu}H \right).
\end{split}
\end{equation}
Also, we assume here that reheating can be described as a perturbative decay of massive oscillating inflaton condensate. This approximation should hold for large enough inflaton masses. The decay width of the inflaton is,
\begin{equation}
\label{Greh}
    \Gamma_{SM}=\frac{\mu^2}{8\pi M}.
\end{equation}
Reheating temperature is related to $\Gamma_{SM}$ as,
\begin{equation}
    T_{reh}=0.3 \,g_{reh}^{1/4}\sqrt{\Gamma_{SM} M_P}.
\end{equation}
Here $g_{reh}$ is the effective number of relativistic degrees of freedom at the moment of reheating ($g_{reh}=106.75$ for the Standard Model). Hereafter we will parameterise reheating by the value of $T_{reh}$ in the example model \eqref{reh}, the coupling $\mu$ can be expressed through $T_{reh}$.

During reheating gravitons will also be emitted due to the effect of bremsstrahlung. The relevant diagrams are presented at Figure 4 in Appendix. The differential decay rate from the first two diagrams is, \footnote{The bremsstrahlung rate grows as $1/k$ at low energies which means that the resummation including soft emission is required for very low frequencies, when $G(k)\sim 1$ \cite{Barker:1969jk,Addazi:2019mjh}. However, in the cases considered in this paper, such low momenta corespond to superhorizon gravitons which we do not include anyway.},
\begin{equation}
   G(k)=A \frac{(m-2 k)^2}{m\, k}+B_{UV}(k), 
\end{equation}
where the expressions for $A$ and $B_{UV}$ are given in the Appendix.

It is worth to mention here that the graviton bremsstrahlung is not affected by any of the extra inflaton couplings to gravity because the diagram (c) is always zero. Thus, the term with $1/\Lambda_4^2$ suppression gives zero impact on the graviton production.



\section{Energy density of gravitons}

The gravitons produced through decay and bremsstrahlung processes remain in the expanding Universe until now. Their energy density contributes to the stochastic gravitational wave (GW) background which can be detected in the future gravitational wave experiments. Hereafter we present an analytic estimate of GW energy density for decay of inflaton to two gravitons and for graviton bremsstrahlung.

{\bf Inflaton decay to two gravitons.} The spectrum of gravitational waves from inflaton decay during early matter domination stage was computed in \cite{Ema:2021fdz,Koshelev:2022wqj}
\begin{equation}
\label{GWs}
    \frac{d\Omega_{GW}}{d \log{k}}=\frac{16 k^4}{m^4}\frac{\rho_{reh}}{\rho_0}\frac{\Gamma_{GW}}{H_{reh}}\frac{1}{\gamma(k)}e^{-\gamma(k)}.
\end{equation}
Here we defined $\rho_{reh}$ and $\rho_0$ to be the energy density at reheating and now, correspondingly, $k$ is the graviton energy now, $H_{reh}$ is the Hubble parameter at reheating, and
\begin{equation}
    \gamma(k)=\left(\left(\frac{g_{reh}}{g_0}\right)^{1/3}\frac{T_{reh}}{T_0}\frac{2 k}{m}\right)^{3/2}.
\end{equation}
 Here $g_0=2 + 7/8\cdot(6\cdot4/11)$, $g_{reh}=106.75$ are the effective number of relativistic degrees of freedom in the SM plasma at reheating and at present time. The GW spectrum has a cutoff in frequency from both IR and UV side because the gravitons were emitted between inflation and reheating. While the UV cutoff is included in \eqref{GWs}, we still need to place the IR cutoff at the frequency 
\begin{equation}
    k_{min}=\frac{m}{2}\frac{a_i}{a_0}=\frac{m}{2}\left(\frac{\rho_{inf}}{\rho_{reh}}\right)^{1/3}\frac{T_0}{T_{reh}}\left(\frac{g_0}{g_{reh}}\right)^{1/3}
\end{equation}
corresponding to the transition from inflation to reheating.  In terms of GW frequency $f=k/(2\pi)$,
\begin{equation}
\label{fmin}
\begin{split}
 f_{min}&=(3.6\times 10^5~{\rm Hz})\times\\
 &\times\frac{m}{10^{13}~{\rm GeV}}\left(\frac{H_{inf}}{10^{14}~{\rm GeV}}\right)^{-1/3}\left(\frac{T_{reh}}{10^{8}~{\rm GeV}}\right)^{1/3}.
\end{split}
\end{equation}
During inflation gravitons were not produced by this mechanism. The form of the GW signal around this frequency is very sensitive to the dynamics of the transition from inflation to reheating. We leave the detailed description of these effects for the future study.

Gravitational wave spectrum is usually expressed in terms of the characteristic strain $h_c(f)$, where $f=k/(2\pi)$. It is connected to the energy density spectrum as follows
\begin{equation}
    h_c(f)=\sqrt{\frac{3 H_0^2 }{\pi f^2}\frac{d\Omega_{GW}}{d f}}.
\end{equation}
We present our results for the inflaton decay in Figure 3, comparing them with the projected sensitivities of the future GW experiments in high frequency range. 

{\bf Graviton bremsstrahlung.} Hereafter we estimate the spectrum of the present gravitational waves from bremsstrahlung effect.
\begin{equation}
    \frac{d\rho_{GW}}{dk}=\int\frac{k dN}{a_0^3}=\int dt\frac{k n_{\phi}(t)a(t)^3}{a_0^3}G(k\frac{a_0}{a(t)})
\end{equation}
The number density $n_{\phi}$ of the inflaton condensate decaying with the rate $\Gamma_{tot}$ can be computed as
\begin{equation}
    n_{\phi}=\frac{\rho_{reh}}{m}\left(\frac{a_{reh}}{a}\right)^3 e^{-\Gamma_{tot} t}.
\end{equation}
It is convenient to introduce dimensionless integration parameter $z(t)=a_{reh}/a(t)$. Switching from the physical time to this variable we obtain
\begin{equation}
\label{domega}
\begin{split}
    &\frac{d\Omega_{GW}}{d \log{k}}=\\
    &\frac{k^2}{m\,H_{reh}}\frac{a_{reh}^2}{a_0^2}\frac{\rho_{reh}}{\rho_0}\int_{z_{min}}^{z_{max}} dz\, G(k z\frac{a_0}{a_{reh}}) z^{-3/2}e^{-2 z^{-3/2}/3}.
    \end{split}
\end{equation}
Here $\Omega_{GW}=\rho_{GW}/\rho_0$, $\rho_0$ is the current energy density of the Universe. The integral in Eq. \eqref{domega} can be performed analytically. We define here the function $F(z)$ emerging after evaluation of the integral,
\begin{widetext}
\begin{equation}
    F_1(z)=\left( \frac{m^2}{E}+2 E\sqrt{z}\right) e^{-2z^{-3/2}/3}-2\left(\frac{2}{3}\right)^{2/3}m\, \Gamma\left(\frac{1}{3},\frac{2}{3}z^{-3/2}\right)-2\left(\frac{2}{3}\right)^{1/3} E\, \Gamma\left(\frac{2}{3},\frac{2}{3}z^{-3/2}\right),
\end{equation}
\end{widetext}
where $\Gamma(x,y)$ is an incomplete gamma-function.

Thus, the GW spectrum can be written as,
\begin{equation}
\label{domega}
    \frac{d\Omega_{GW}}{d \log{k}}=\frac{k^2}{M\,H_{reh}}\frac{a_{reh}^2}{a_0^2}\frac{\rho_{reh}}{\rho_0} A (F_1(z_{max})-F_1(z_{min}))
\end{equation}
\begin{equation}
z=\frac{a_{reh}}{a}, ~~\frac{a_{reh}}{a_0}=\frac{T_0}{T_{reh}}\left(\frac{g_0}{g_{reh}}\right)^{1/3}.
\end{equation}

\begin{figure}[htb]
\label{bounds}
    \begin{center}
  \includegraphics[scale=.5]{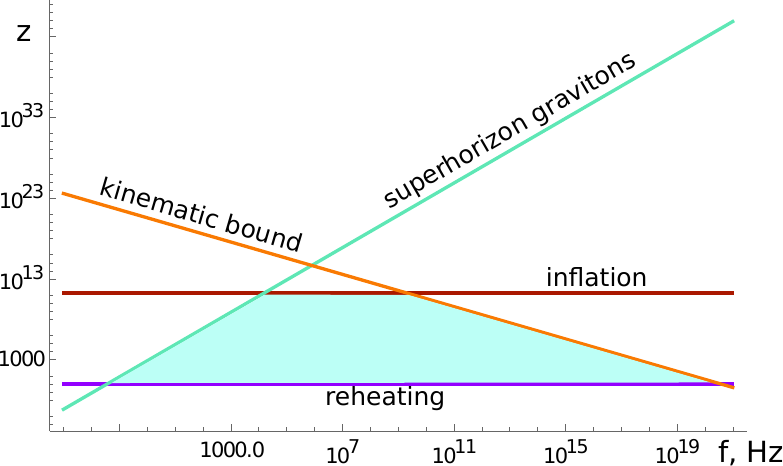}
  \end{center}
\caption{Illustration of integration limits for $z$ for $T_{reh}=10^8$ GeV and inflation Hubble scale $H_{inf}=10^{14}$ GeV.}
\end{figure} 

What are the limits of the integral, $z_{min}$ and $z_{max}$? They depend on the frequency of the graviton which we probe now, confining the period when gravitons contributing to the probed mode $k$ were emitted in the early Universe. There are three relevant bounds on $z$.
\begin{itemize}
    \item The gravitons were emitted between inflation and reheating, $1<z<a_{inf}/a_{reh}$ where $a_{inf}$ is the scale factor at the end of inflation. Thus, if $\rho_i$ and $H_{inf}$ are the inflaton energy density and Hubble scale at inflation,
    \begin{equation}
    z<z_1=\left(\frac{\rho_{inf}}{\rho_{reh}}\right)^{1/3}=\left(\frac{3 H_{inf}^2 M_P^2}{g_{reh} T_{reh}^4\,\pi^2/30}\right)^{1/3}.
    \end{equation}
    \item We can use the described approach only for the subhorizon gravitons, $k\gg a(z) H(z)$. Assuming matter dominated stage preceding reheating, this implies the condition
    \begin{equation}
        z<z_2=\frac{k^2 a_0^2}{a_{reh}^2 H_{reh}^2}.
    \end{equation}
    This condition is the reason for the sharp IR cutoff of the GW spectra in Figure 4. As bremsstrahlung is a process localized in space, production of superhorizon gravitons must be strongly suppressed because it violates causality. Leaving the accurate computation of this process for $z\approx z_2$ for future study, in this work, we simply cut the integral at $z=z_2$ which should be still a good approximation.
    
    \item There is a kinematic upper bound $m/2$ on the comoving energy of the produced graviton which implies
    \begin{equation}
        z<z_3=\frac{m}{2 k}\frac{a_0}{a_{reh}}.
    \end{equation}
\end{itemize}
The listed conditions provide an upper bound $z_{max}={\rm Min}\,(z_1,z_2,z_3)$ for the integral \eqref{domega}. The lower limit is determined by the decay of inflaton particles and reheating, $z_{min}=1$. We visualize the described bounds in Figure 1 for $T_{reh}=10^8$ GeV and show that all of them are relevant for the computation\footnote{The recent paper \cite{2301.11345} presents the results for GW production from bremsstrahlung which partially coincide with obtained in this work for $\Lambda_{UV}=M_P$. The main difference is that they do not take into account IR cutoff related to superhorizon modes, thus, they overestimated GW spectrum at low frequencies.}. 

\section{Results and conclusions}

Graviton production after inflation leads to stochastic GW background contribution at very high frequencies. Detection methods for GWs is currently the actively developing area of research \cite{1607.08697,     1807.09495,1512.02076,astro-ph/0108011,Punturo:2010zz,1702.00786}. We present the results of our computations in Figures 2 and 3 for GW spectra plotted over the projected sensitivities of various high frequency GW detection proposals, including \cite{2010.13157,1611.05560,1410.2334,2112.11465,2203.15668}, see also the Living review \cite{2011.12414}. 

As a main conclusion, let us stress here that the contribution of even Planck suppressed inflaton and matter couplings to gravity can lead to potentially observable outcome in high frequency GWs and in dark radiation measurements \cite{2004.11392,1508.02393,2004.11396}. Thus, even if the theory is valid until the Planck scale, it does not mean that the higher derivatives can be always safely omitted. Needless to say, if the unitarity breaking scale for the model describing inflation and reheating is lower than the Planck scale, the described effects are getting stronger. In particular, the bound on $\Delta N_{eff}$ leads to several reheating-dependent constraints on the unitarity breaking scale. Although the exact bounds are model dependent, it can be observed from the plots that having $\Lambda_{UV}\lesssim 10^{-3} M_P$ can lead to a contradiction with observations. 

The methods and results of this work can be straightforwardly generalised to the other couplings between matter (including fermions and vectors \cite{2301.11345,2211.10433}) and gravity, as well as for the other mechanisms of reheating. Even if the inflaton has $\phi\rightarrow -\phi$ symmetry forbidding some operators studied here, the gravitons still would be produced during reheating and preheating stages, however, the computations are much more complicated and may require non-perturbative methods in this case. The results obtained on a lattice show the high frequency GW signal from preheating as a generic feature of this epoch \cite{hep-ph/9701423,astro-ph/0601617,gr-qc/9909001,1707.04533,2206.14721,2312.15056}.

\begin{figure}[htb]
\label{decay}
    \begin{center}
  \includegraphics[scale=.5]{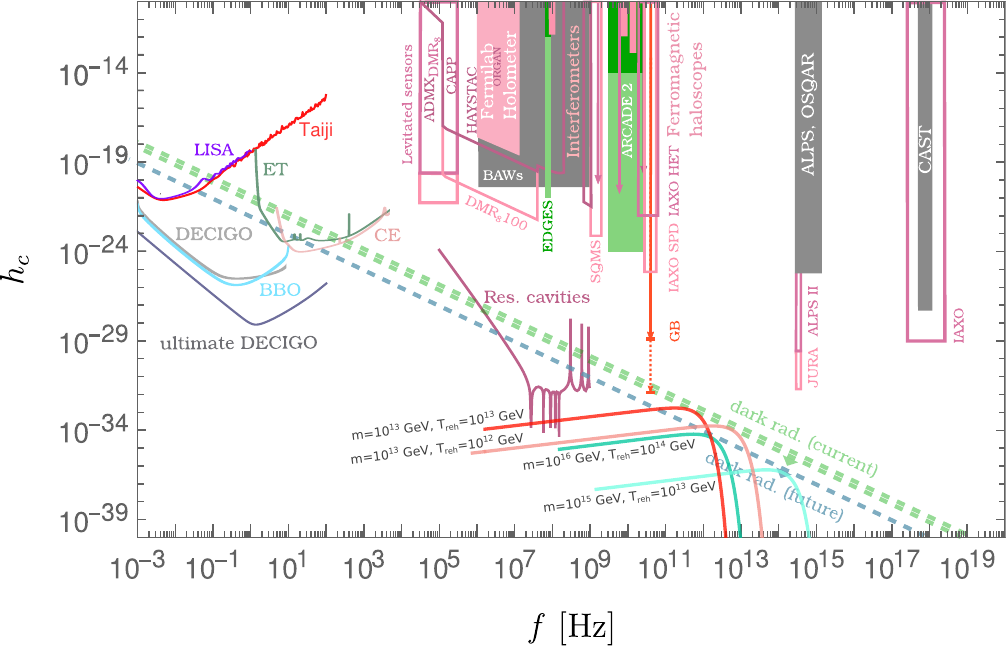}
  \end{center}
\caption{Gravitational wave spectra from inflaton decay. The mass and temperature is given in GeV. Red curves saturate the current dark radiation bound while the blue curves correspond to the choice $\Gamma_{GW}=0.01\Gamma_{SM}$. For all these plots $\Lambda_{UV}\gtrsim 10 \,m$ so the computations are still within the regime of validity of EFT.}
\end{figure}   

\begin{figure}[htb]
\label{br}
    \begin{center}
  \includegraphics[scale=.5]{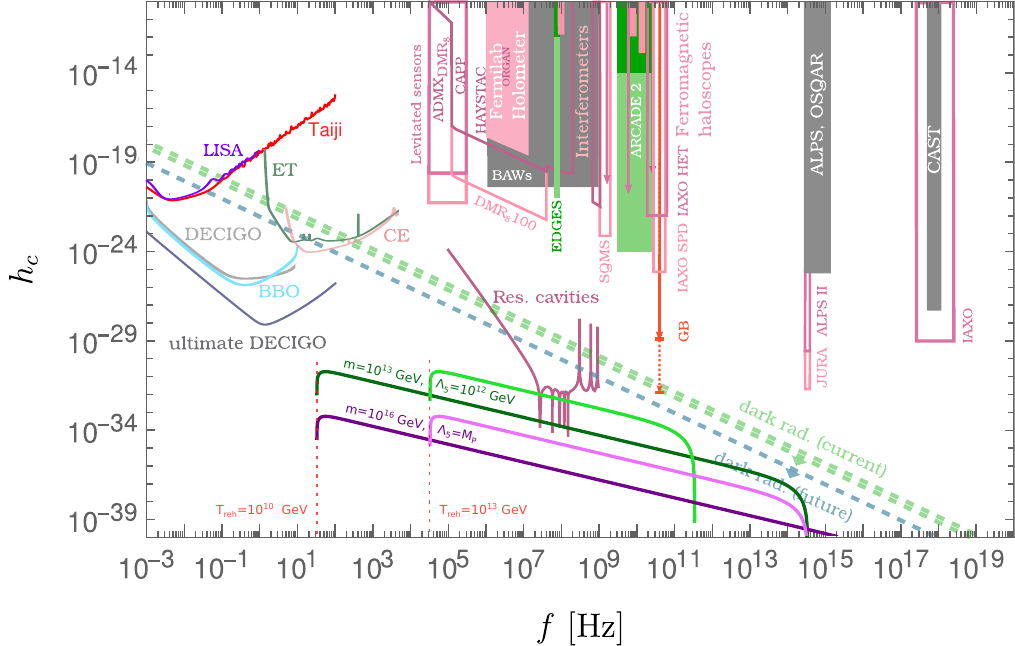}
  \end{center}
\caption{Gravitational wave spectra from graviton bremsstrahlung during the inflaton decay. The mass, scale $\Lambda_5$ and temperature are given in GeV. The plots shown here illustrate the two cases: heavy inflaton ($10^{16}$ GeV) with Planck suppressed couplings only (violet colours), and lighter inflaton ($m=10^{13}$ GeV) with $\Lambda_5=10^{12}$ GeV corresponding to the lower unitarity breaking scale $\Lambda_{UV}\sim 10^{15}$ GeV (green colours). Both cases are within the reach of resonant cavities GW experiments proposed for the future.}
\end{figure}

Our study of graviton production in the early Universe is incomplete from the side of reheating, as we do not compute graviton production in hot plasma. There are a lot of recent papers presenting state-of-art computations of GWs from the hot stage \cite{1504.02569,2004.11392,2011.04731,1801.04268,2203.00621,1810.04975,1905.08510,Klose:2022knn,Ghiglieri:2022rfp,Ghiglieri:2024ghm}, where effects of interactions in a dense plasma are crucial and lead to an upper bound on $\Omega_{GW}$ \cite{2312.13855}. As the graviton production we computed here happens when there are not so many matter particles yet, it is safe to neglect these effects in the current work. 

\begin{acknowledgments}
The author is grateful to Marco Drewes for the fruitful discussion and sharing his concerns about the limitations of the approach. The athour also thanks Andreas Ringwald and Simona Procacci for their feedback during the CERN TH insitute 'Ultra-high frequency gravitational waves: where to next?'. The work of AT at the beginning was supported by STFC grant ST/T000791/1. At the final stage AT is supported in part by the National Natural Science Foundation of China (NSFC) under Grant No. 12147103.
\end{acknowledgments}

\section*{Appendix}

Here we give the definitions and explicit expressions for the graviton bremsstrahlung rate in the model where the reheating vertex and decay width is described by Eq. \eqref{reh} and \eqref{Greh}.
\begin{equation}
   G(k)= \frac{\partial \Gamma}{\partial k}=\frac{1}{(2\pi)^3}\frac{1}{8 m}\int_{q_{min}}^{q_{max}}|{\cal M}(q,k)|^2 dq.
\end{equation}
The diagrams (c) and (d) do not contribute to the graviton emission for kinematical reasons and due to orthogonality of graviton polarisation tensor. The remaining contributions give
\begin{equation}
   G(k)=A \frac{(m-2 k)^2}{m\, k}+B_{UV}(k), 
\end{equation}
where
\begin{equation}
    A=\frac{1}{64 \pi^3}\frac{\mu^2}{2 M_p^2}\left(\frac{m^2}{\Lambda_5^2}+1\right)^2,
\end{equation}
and the part of bremsstrahlung rate which is subdominant for $k\ll m$ is
\begin{equation}
  B_{UV}(k)=\frac{1}{64 \pi^3}\frac{\mu^2}{2 M_p^2}\frac{2 (m-2 k)^2}{15\Lambda_5^2}\left(\frac{m\,(7 k-10 m)}{\Lambda_5^2}-10\right).
\end{equation}
This part does not appear from the graviton coupling to the kinetic term, so it is relevant only if quantum gravity scale is significantly lower than the Planck mass.

\begin{widetext}

\begin{figure}[htb]
\label{diagrams}
    \begin{center}
  \includegraphics[scale=.28]{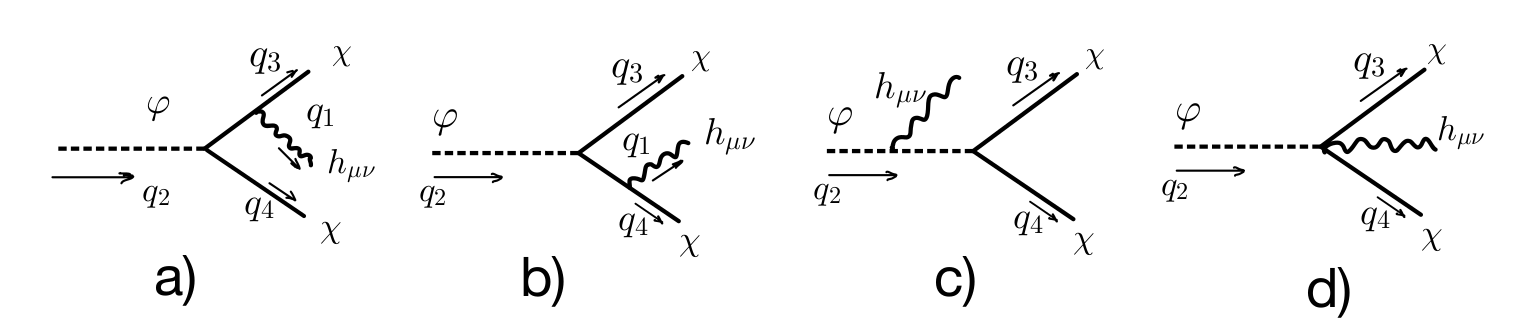}
  \end{center}
\caption{Diagrams contributing to the graviton bremsstrahlung. Here $\chi$ stands for each scalar degree of freedom (in the case of the SM Higgs there are four scalars, hence the extra factor 4 in the bremsstrahlung rate). The momenta can be chosen in the rest frame of the inflaton, $q_2=(m,0,0,0)$, $q_1=(k,0,0,k)$, $q_3=(q,q\sin{\theta},0,q\cos{\theta})$, $q_4=(m-q-k,-q\sin{\theta},0,-q\cos{\theta}-k)$, where $\theta$ is the angle between spatial momenta of decay products. We assume the decay products are massless which is a very good approximation during reheating, as the temperature is much larger than Higgs mass. Diagrams c) and d) give zero contribution because of the properties of graviton polarization tensors.}
\end{figure}    

\end{widetext}


\end{document}